\def\grs{GRS~1915+105}
\def\X1550{XTE~J$1550-564$}
\def\J1655{GRO~J$1655-40$}

\def\igr{IGR~J19294+1816}
\def\ergcms{erg~cm$^{-2}$~s$^{-1}$ }
\def\integ{{\it{INTEGRAL}}}
\def\xte{{\it{RXTE}}}
\def\swift{{\it{Swift}}}
\def\nh{N$_\mathrm{H}$}
\documentclass[traditabstract]{aa}
\usepackage{natbib,epsfig, epsf,color,txfonts}
\definecolor{red}{rgb}{0.7,0,0}
\definecolor{blue}{rgb}{0,0,0.7}
\def\correc#1{{#1}}
\usepackage{amssymb}

\begin{document}

\titlerunning{IGR J19294+1816 as seen with 
{\em INTEGRAL}, {\it RXTE}, and {\it Swift}}
\authorrunning{Rodriguez et al.}
\title{The nature of the X-ray binary \object{IGR J19294+1816} from 
{\em INTEGRAL}, {\em RXTE}, and {\em Swift} observations}

\author{J. Rodriguez\inst{1}, J.A. Tomsick\inst{2}, A. Bodaghee\inst{2}, J.-A. Zurita Heras\inst{1},  
S. Chaty\inst{1}, A. Paizis\inst{3}, S. Corbel\inst{1}}

\offprints{J. Rodriguez \email{jrodriguez@cea.fr}}

\institute{Laboratoire AIM, CEA/DSM - CNRS - Universit\'e Paris Diderot (UMR 7158), CEA Saclay, DSM/IRFU/SAp, F-91191 Gif-sur-Yvette, France \and
Space Science Laboratory, 7 Gauss Way, University of California, Berkeley, C1 94720-7450, USA \and
 INAF-IASF, Sezione di Milano, via Bassini 15, 20133 Milano, Italy}


\abstract{We report the results of a high-energy multi-instrumental campaign with 
\integ, \xte, and \swift\ of the recently discovered \integ\ source \igr. The \swift/XRT
data allow us to refine the position of the source to 
RA$_{\mathrm{J}2000}$= 19$^h$ 29$^m$ 55.9$^s$ 
Dec$_{\mathrm{J}2000}$=+18$^\circ$ 18\arcmin\ 38.4\arcsec\ ($\pm 3.5$\arcsec), 
which in turn permits us to identify a candidate infrared counterpart. The \swift\ and \xte\ spectra are well
fitted with absorbed power laws with hard ($\Gamma \sim 1$) photon indices.  
During the longest \swift\ observation, we obtained 
evidence of absorption in true excess to the Galactic value, which 
may indicate some intrinsic absorption in this source. We detected 
a strong (P=40\%) pulsation 
at $12.43781 (\pm0.00003)$~s that we interpret as the spin period of a pulsar. All these 
results, coupled with the possible 117 day orbital period, point to \igr\ being an HMXB with 
a Be companion star. However, while the long-term \integ/IBIS/ISGRI  18--40 keV 
light curve shows that the source spends most of its time in an undetectable state, we 
detect occurrences of short ($\sim 2000-3000$~s) and intense flares that are more 
typical of supergiant  fast X-ray transients.  We therefore cannot make firm conclusions on the type of system, and we discuss the possible implications 
of \igr\ being an SFXT. }

\maketitle

\section{Introduction}
\indent The INTErnational Gamma-Ray Astrophysics Laboratory (\integ) has permitted 
a large number of new sources to be discovered. Amongst the $\sim250$ new 
sources\footnote{see http://isdc.unige.ch/$\sim$rodrigue/html/igrsources.html for an up to date
list of all sources and their properties.}, \integ\ has unveiled two new or poorly-known
types of high mass X-ray binaries (HMXB): 
the very absorbed supergiant HMXBs, and the supergiant fast X-ray 
transients (SFXT). While all types of HMXBs (including those hosting a Be star, Be-HMXBs)
are powered by accretion of material by a compact object, understanding the nature of 
a system is very important for studying the evolutionary paths of source populations, and more globally, the evolution of the Galaxy in terms of its source
content.\\ 
\indent \igr\ was discovered by \citet{2009ATel.1997....1T} who reported activity of 
this source
seen with \integ\ during the monitoring campaign of  \grs\ \citep[e.g.][]{rodrigue08_1915a}. Archival \swift\ data of a source named \object{Swift J1929.8+1818} allowed us to give a 
refined X-ray position of RA$_{\mathrm{J}2000}$= 19$^h$ 29$^m$ 55.9$^s$ and Dec$_{\mathrm{J}2000}$=+18$^\circ$ 18\arcmin\ 39\arcsec\ ($\pm 3.5$\arcsec) 
which we used to locate a possible infrared counterpart \citep{2009ATel.1998....1R}. 
We also suggested that the \swift\ and \integ\ sources are the same and that activity from \igr\ 
had been seen with \swift\ in the past.  The temporal analysis of the XRT light curve showed 
a possible pulsation at 12.4~s \citep{2009ATel.1998....1R}. 
Analysis of \swift/BAT archival data revealed that the source had 
been detected on previous occasions with a periodicity at 117.2 days 
\citep{2009ATel.2008....1C} that we interpreted as the orbital period of the system. \\
\indent Soon after the discovery of the source with \integ, we triggered our accepted \xte\ and
\swift\ programmes for follow-up observations of new \integ\ sources (PI Rodriguez). A preliminary 
analysis of the real-time \xte\ data allowed us to confirm the pulsations
in the signal from the source \citep{2009ATel.2002....1S} at a barycentred period of 
12.44~s indicating 
that this object hosts an accreting X-ray pulsar. Here we report the detailed analysis of the \integ, 
\swift, and \xte\ observations. We begin this paper by detailing the procedures employed for the 
data reduction. We then describe the results (refined position, X-ray spectral, 
and temporal analyses) in Sects. 3, 4, 5, and 6, and discuss them in Sect. 7.

\section{Observations and data reduction}
The log of the \swift\ and \xte\ observations of \igr\ is given in Table~\ref{tab:log}.  
The {\tt LHEASOFT v6.6.2} suite was used to reduce the \swift\ and \xte\ data.
The \xte/PCA data were reduced in a standard way, restricting it to the times when 
the elevation angle above the Earth was greater than 10$^\circ$, and when the offset pointing was less than 
0.02$^\circ$. We accumulated spectra from the top layer of proportional counter unit 2 (PCU 2), 
the only one that was active during all the observations, and therefore time-filtered the 
data for PCU 2 breakdown. The background was estimated using the faint model. 
In addition, 
we produced event files from {\tt {Good Xenon}} data with {\tt {make\_se}}, from which we
 extracted $2^{-12}$~s ($\sim 250$~$\mu$s) light curves \correc{between 2 and 20.2 keV 
(absolute channels 5 to 48 in order to limit the instrumental background at low and high 
energies)} with {\tt {seextrct}}. Light 
curves with 1-s time resolution were obtained in the 2--60 keV range from standard 1 data. These light curves
were then corrected for the background and used for the temporal analysis.\\
\indent The \swift/XRT data were first processed with the {\tt{xrtpipeline v0.12.2}} tool to 
obtain standardly filtered level 2 event files. Images, light curves, and spectra were extracted 
within {\tt{XSELECT V2.4a}}. Source light curves and spectra were extracted from a 
circular region of 20 pixels radius centred on the most accurate source position available. We tested different 
regions of extraction for the background light curves and spectra, and obtained no significant 
differences. For the remainder of the analysis, the background region is a circular region of 60 
pixels radius centred at RA=19$^h$ 29$^m$ 41.8$^s$ and 
Dec=+18$^\circ$ 21\arcmin\ 11.5\arcsec. The ancillary 
response file was obtained with {\tt xrtmkarf 0.5.6} after taking the exposure map 
into account \citep[see][for more details]{rodrigue08_igr}.\\
\begin{table}
\caption{Journal of the \swift\ and \xte\ dedicated observations. Obs. S3 and X5 were
simultaneous with an \integ\ observation.}
\begin{tabular}{c c c c}
\hline
\hline
Satellite & Date start & Exposure time & Name\\
             & (MJD)    & (s)\\
\hline
\swift &54443.051 & 7959 & S1 \\
\swift & 54447.013 & 3457 &S2 \\
\xte & 54921.315 & 2600 & X1\\
\xte & 54921.708 & 3344 & X2\\
\xte & 54922.813 & 3344 & X3\\
\xte & 54923.730 & 1440 & X4\\
\swift &54925.814 &2565 & S3\\
\xte & 54925.826 & 3360& X5\\
\hline
\end{tabular}
\label{tab:log}
\end{table}
\indent We also analysed  {\it all} \integ\ public pointings 
aimed at less than 10$^\circ$ from the position of \igr\ plus all private data 
of our monitoring 
campaign on \grs.  The  {\tt Offline Scientific Analysis Software (OSA) v7.0} was used for 
the \integ\ data reduction.  We further filtered the list of science windows (SCW) 
to those where the 
good exposure time of ISGRI was greater than 1000~s. After omitting the bad pointings (e.g. 
first or last pointings before the instruments are switched off due to passage through the 
radiation belts) this resulted in a total of 1476
good SCWs. These pointings were blindly searched 
for the presence of the source (see \citet{rodrigue08_1915a} for the procedure of 
data reduction and source searching). An updated catalogue containing \igr\ with the most precise \swift\ position was given as an input to the software.  Mosaics were accumulated on 
a revolution basis, but this results in mosaics of very different exposure times. 
Note that even in revolution 788, where we confirm the detection reported by 
\citet{2009ATel.1997....1T}, the source is too dim for further spectral analysis. \\

\section{\igr\ as seen with \integ}
\begin{figure}
\epsfig{file=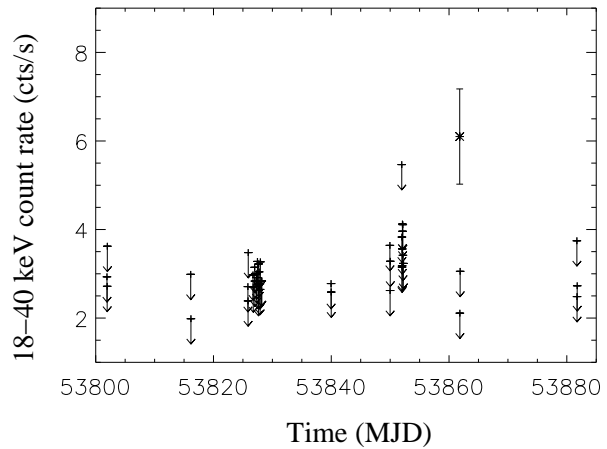, width=\columnwidth}
\caption{\integ/IBIS 18--40 keV light curve of \igr\ showing the flare on MJD 53861.}
\label{fig:igrlite}
\end{figure}

\igr\ was discovered on March 27, 2009 \citep[revolution 788, MJD 54917][]{2009ATel.1997....1T}.
A re-analysis of the consolidated data allowed us to confirm a source detection 
during this observation.  The source was detected at 10.7 $\sigma$
between 18 and 40 keV, with a flux of about 17 mCrab\footnote{The count 
rate to Crab conversion is based on an analysis of the Crab observation 
performed during rev. 774, with 1 Crab$_{18-40\mathrm{keV}} = 181.1$~cts/s.}.  
A re-analysis of the previous observation on March 21, 2009 (rev. 786, MJD 54911), 
led to a detection in the 18--40~keV energy range with a significance of 
4.5~$\sigma$ and a flux of 7.2~mCrab. \igr\ was not detected on 
April 4, 2009 (rev 790, MJD 54925) with a 18--40 keV 3-$\sigma$ upper limit 
of about 5 mCrab.  The source was not detected in any of the 
mosaics from the previous revolution. However, given the very different exposure times we 
used in building the mosaics and the periodicity of the source activity seen 
with \swift/BAT \citep{2009ATel.2008....1C}, this may simply indicate that the periods of 
activity of the source are short. \\
\indent Examining these data on a SCW basis shows that the source was detected on a few occurrences. 
In particular, it was clearly detected during an observation in rev. 435 (MJD 53861) and during 
another pointing in
rev. 482 (MJD 54004). In the former, \igr\ was detected in SCW 9 (19.17--19.77 h UTC) with 
a 18--40 keV flux of 33.7 mCrab. It was not detected in the following 
SCW (19.85--20.50 h) with a 3-$\sigma$ upper limit of 11.6 mCrab. No pointings are 
available before SCW 9. The 18--40 keV light curve of this epoch is reported in Fig.~\ref{fig:igrlite}.
In rev. 482, the source was detected in SCW 51 (1.05--2.03 h UTC) 
at a 18--40 keV flux of 22 mCrab (5.4 $\sigma$). It was not detected in the 
previous and following SCWs with respective 3-$\sigma$ upper limits of 
11.0 and 13.9 mCrab. 
\section{Refining the X-ray position, and the search for counterparts}
\begin{figure}
\epsfig{file=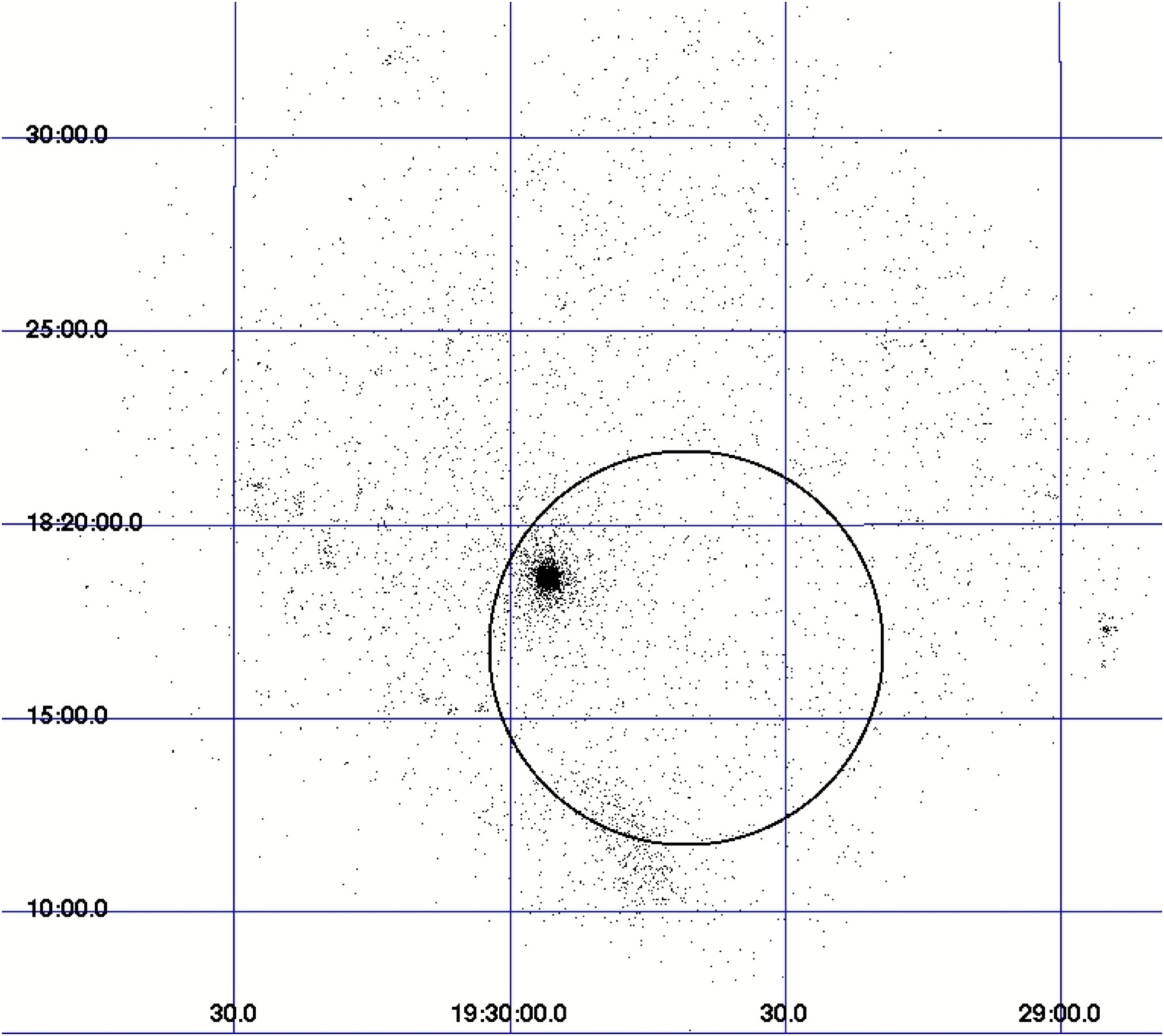,width=\columnwidth}
\caption{\swift/XRT mosaic of the field around \igr. The circle represents the 
error box from \integ.}
\label{fig:x-ray}
\end{figure}

\begin{figure*}
\centering
\epsfig{file=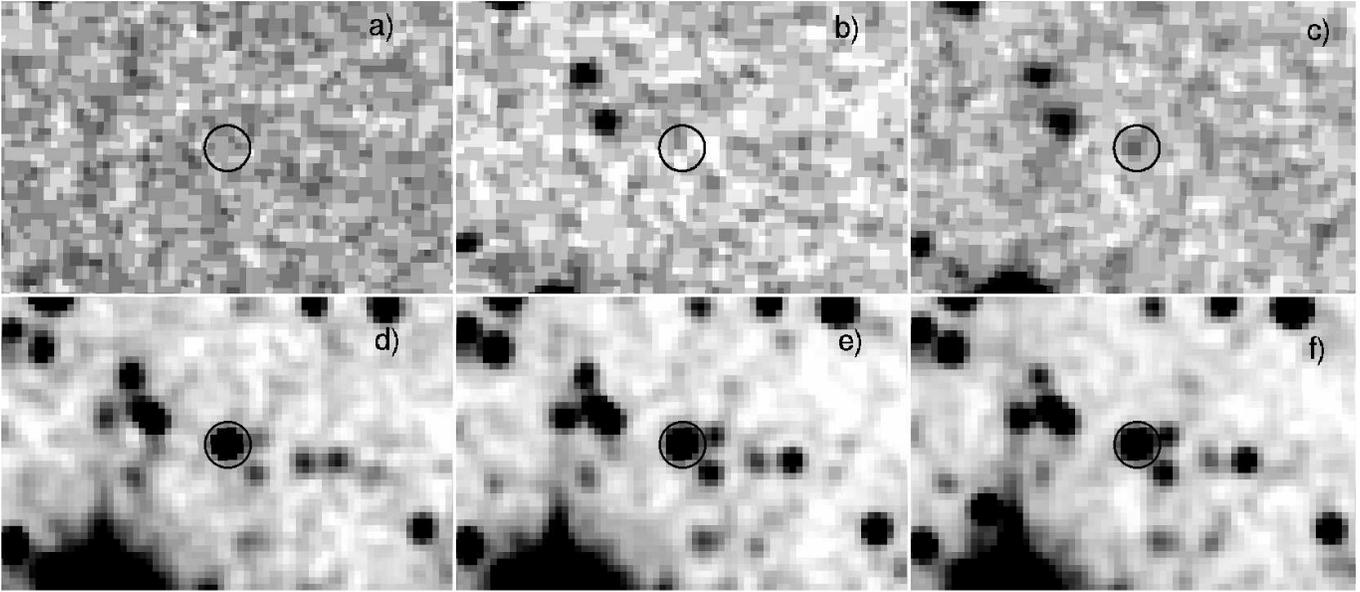,width=18cm}
\caption{1.14\arcmin$\times$0.73\arcmin\ POSS II optical (top) and 2MASS infrared (bottom) 
images of the field around \igr. The black circle is the \swift\ error box. a) `` Blue'', b)
``Red'', 
c) ``infrared'', d) J filter e) H filter, f) K$_{\mathrm s}$ filter.}
\label{fig:charts}
\end{figure*}

We combined all  \swift\ observations to create a mosaic image of the field. 
As shown in Fig.~\ref{fig:x-ray}, a single bright X-ray source can be seen 
within the $\sim5$\arcmin\ 90\% error circle of the \integ\ position given in 
\citet{2009ATel.1997....1T}. This source lies at (as obtained with {\tt xrtcentroid})
RA$_{\mathrm{J}2000}$=19$^h$ 29$^m$ 55.9$^s$ and Dec$_{\mathrm{J}2000}$=+18$^\circ$ 18\arcmin\  38.4\arcsec\ 
with an uncertainty of 3.5\arcsec. Note that this is slightly different from the preliminary 
position given in \citet{2009ATel.1998....1R}, but the two positions are clearly compatible. 
Given the addition, here, of an additional \swift\ pointing to obtain these coordinates, we consider this 
position as the most precise one. \\
\indent We searched for counterparts in catalogues available online such as the 2 Microns All Sky Survey point source and extended source catalogues\footnote{http://www.ipac.caltech.edu/2mass/} 
\citep[2MASS and 2MASX,][]{skrutskie06}, the National Radio Astronomy Observatory VLA Sky
Survey \citep[NVSS,][]{condon98}, and the catalogue of the US Naval Observatory (USNO\footnote{http://www.usno.navy.mil/USNO/astrometry/optical-IR-prod/icas}). We also made use 
of images from the Second Palomar Observatory Sky Survey obtained through the STScI
Digitized Sky Survey (DSS\footnote{http://archive.stsci.edu/cgi-bin/dss$\_$form}). 
 Finding charts from 2MASS IR and DSS are shown in Fig.~\ref{fig:charts}.

A single infrared counterpart is listed in the 2MASS catalogue: 
\object{2MASS~J19295591+1818382} has magnitudes J=14.56 $\pm$0.03, 
H=12.99 $\pm$0.02, and K$_{\mathrm s}$=12.11 $\pm$0.02. No optical or radio counterpart was 
found in any of the other catalogues, although a faint object might 
be present in the ``infrared'' (7000--9700\AA) POSS II image (Fig.~\ref{fig:charts}).
The infrared magnitudes were dereddened using A$_\lambda=C(\lambda) \times$A$_V$, 
with $C(\lambda)$  the wavelength dependent coefficients obtained by \citet{cardelli89} 
\citep[see Table 4 in][for the values adopted]{rahoui08}, and 
A$_V$=5.6$\times 10^{22} \times$ \nh\ \citep{predehl95}.
The value of the total absorbing column density along the line of sight 
was obtained from the LAB survey of Galactic HI \citep{kalberla05}, and the CO survey 
of \citet{dame01} for the value of molecular H$_2$. 
Therefore, \nh$=N_{\mathrm {HI}}+2N_{\mathrm H_2}=1.23 + 
2\times0.75=2.73 \times 10^{22}$~cm$^{-2}$. 
We also used the value of \nh\ obtained through the X-ray spectral fits to the data, 
\nh$_{\mathrm{X,bbody}}$=2.0$\times 10^{22}$ cm$^{-2}$, and  \nh$_{\mathrm{X,po}}$=
4.0$\times 10^{22}$ cm$^{-2}$ (see Sec. \ref{sec:spectral}).  
The dereddened magnitudes are reported in Table~\ref{tab:mag}.
\begin{table}
\caption{Infrared magnitudes dereddened with different values of 
the absorption.}
\label{tab:mag}
\begin{tabular}{ccccccc}
\hline\hline
\nh & A$_V$ &J$_{\mathrm {der.}}$ & H$_{\mathrm{der.}}$ & K$_{\mathrm{s,der.}}$ & J-H & H-K$_{\mathrm s}$\\
$\times 10^{22}$ cm$^{-2}$ & & & & & \\
\hline
2.0   & 11.20 & 11.32 & 11.04 & 10.86  & 0.28      & 0.18\\
2.73 & 15.29 & 10.14 & 10.32 & 10.40 & $-$0.18 & $-$0.08\\
4.0   & 22.40 & 8.08   &  9.09  & 9.53   & $-$1.01 & $-$0.44\\
\hline
\end{tabular}
\end{table}
\section{X-ray spectral analysis}
\label{sec:spectral}
\subsection{Correction of the Galactic ridge emission in the PCA spectra}
Since \igr\ is faint and located in the Galactic plane (its Galactic coordinates are 
$l=53.5400^\circ$, $b=0.1150^\circ$), the Galactic ridge can contribute
a significant amount of the flux collected by the PCA. We estimated 
the level of emission using the results from \citet{valinia98}, and first 
 corrected the spectra from the Galactic ridge 
emission in a similar way as done for IGR J19140+0951 by \citet{lionel08}, i.e.
adding the Galactic ridge spectrum to the instrumental background spectrum. \\
\indent A simultaneous fit to the S3 and X5 spectra showed a large offset in 
the normalisations of the spectra: when including a constant in the fit and freezing 
it to 1 for the S3 spectrum it resulted in a low value of $\sim$0.45 for the 
X5 spectrum. This might indicate that either the source varied during
the non-simultaneous part of these observations, or that we over-corrected the emission 
of the Galactic ridge in the PCA spectrum. We consider this last hypothesis as the most 
likely given that the spectrum of the Galactic ridge obtained from \citet{valinia98} 
is an average spectrum covering the central regions of the Galaxy, 
i.e. $-45^\circ < l < 45 ^\circ$ and $-1.5^\circ < b < 1.5^\circ$, and that \igr\ is 
slightly outside this region. Although the source is located in the Galactic ridge, we can 
expect local variations in the ridge, and/or slight changes (lower intensity for 
example) at high longitude \citep[as also mentioned in][]{valinia98}.  \\
\indent In a second run, we did not correct the X5 spectrum for the contribution of 
the Galactic ridge, and we did not include a normalisation constant 
between the S3 and X5 spectra (we therefore made the assumption that
\swift\ and \xte\ are perfectly cross calibrated). The spectra were fit with an absorbed power law and the best-fitting model for the Galactic ridge spectrum obtained by \citet{valinia98}. A multiplicative constant precedes this spectral component whose parameters are frozen. The spectral model is therefore {\tt {phabs*powerlaw+constant*(wabs*(raymond+powerlaw))}} in the {\tt XSPEC} terminology. We initially fitted the S3 spectrum alone (with the constant 
fixed at 0) to obtain the parameters of the source. Then, we added the X5 spectrum
and only left the normalisation constant to the X5 data free to vary while it was
held at 0 for the S3 data. Therefore, we assume that the average spectral parameters obtained in the $-45^\circ < l < 45 ^\circ$ are still valid at the position of \igr\ , and that only the overall flux of the ridge emission changes.
The most precise value obtained for the constant is 0.85$\pm0.04$ (at 90\% confidence). In subsequent analyses, we used this value to correct all PCA spectra according to: 
\begin{equation}
{\mathrm{BGD_{tot}}={\mathrm{BGD_{instr}}}+0.85\times {\mathrm{Ridge}}} \label{eq:bgd}
\end{equation}

\subsection{Results of the spectral analysis}
The \swift/XRT spectra were rebinned so as to have a minimum of 20 cts per channel, except for the channels below about 1.2~keV that were grouped in 2 bins: one below 0.2 keV that was not 
considered further for the spectral fitting, and one between 0.2--1.2 keV to obtain an additional 
 spectral point to better constrain the absorption column density \nh. \correc{The \swift\ 
spectra were fitted between $\sim0.8$ and $\sim 8$~keV, while the \xte\ spectra were 
fitted between $3$ and $25$ keV.}
We fitted each individual spectrum alone in {\tt XSPEC 11.3.2ag}, except the spectra from
observations S3 and X5 that were fitted simultaneously. Different spectral models were 
used to fit the spectra. The \swift\ spectra are well fitted 
using either an absorbed blackbody or an absorbed power law. Only the latter model
provides a good fit to the \xte\ spectrum, and joint \swift\ and \xte\ spectra. 
Considering the detection at energies greater than 
20 keV with \integ\ and \swift\ , this suggests that the true origin of the emission is probably not from a blackbody. However, we report the parameters of the blackbody 
together with those obtained with the power-law model in Table~\ref{tab:spectral}. 
Observations S3 and X5 were simultaneous with an \integ\ observation, 
but the source was not detected by the latter satellite with an 18--40 keV 3-$\sigma$ 
upper limit of 5~mCrab with IBIS.

\begin{table*}
\centering
\caption{Results of the X-ray spectral fitting. The errors are at the 90\% confidence level.}
\begin{tabular}{l l l l l l}
\hline
\hline
Obs & \nh & $\Gamma$ &  kT$_{bb}$ & F$_{2-10, unabs.}$  & $\chi^2_\nu$$^\star$ \\ 
    & ($\times10^{22}$ cm$^{-2}$) &  & (keV) & (\ergcms) & (DOF$^\dagger$) \\
\hline
S1  & 4.0$_{-0.8}^{+0.9}$ & 0.9$\pm0.2$ & &3.6 $\times 10^{-11}$ & 1.0 (66)\\
    & 2.0$\pm0.4$ &  & 2.1$\pm0.2$ & 3.0 $\times 10^{-11}$ & 0.86 (66)\\
S2  & 3.4$_{-0.8}^{+1.0}$ & 1.0$\pm0.3$ & & 3.8 $\times 10^{-11}$ & 0.78 (34)\\
    & 1.8$\pm0.5$ & & 1.8$_{-0.2}^{+0.3}$ & 3.0 $\times 10^{-11}$ & 0.74 (34)\\
X1  & 3.4 (Frozen) & 1.17$_{-0.11}^{+0.14}$ & &2.65 $\times 10^{-11}$ & 0.94 (25)\\
X2  & 3.4 (Frozen) & 1.0$_{-0.2}^{+0.3}$ & & 0.9 $\times 10^{-11}$ & 0.49 (10)\\ 
X3  & 3.4 (Frozen) & 1.2$\pm0.1$ & &2.6$\times 10^{-11}$  &0.91 (25) \\
X4  & 3.4 (Frozen) & 1.1$\pm0.4$ & &1.1$\times 10^{-11}$  & 0.93 (12)\\
S3+X5 & 3.4$_{-1.6}^{+1.5}$ & 1.2$_{-0.4}^{+0.3}$ & & 0.74$\times 10^{-11}$ & 0.86 (8)\\
\hline
\end{tabular}
\begin{list}{}{}
\item[$^\star$]Reduced $\chi^2$.
\item[$^\dagger$]Degrees of freedom.
\end{list}
\label{tab:spectral}
\end{table*}

Since the PCA is well calibrated only above 3~keV, it is not suitable for proper 
measures of \nh, especially in such a dim source. In fact, leaving all parameters 
free to vary during the spectral fits leads to poorly-constrained values for this
parameter (\nh$<5.4\times 10^{22}$~cm$^{-2}$ in Obs. X1). We therefore froze the value of \nh\
to the value returned by the fit to the closest \swift\ data (Obs. S3 which, simultaneously fitted
with X5, leads to $3.4 \times 10^{22}$~cm$^{-2}$ for a power-law model).  

\section{X-ray temporal analysis}
\begin{figure}
\epsfig{file=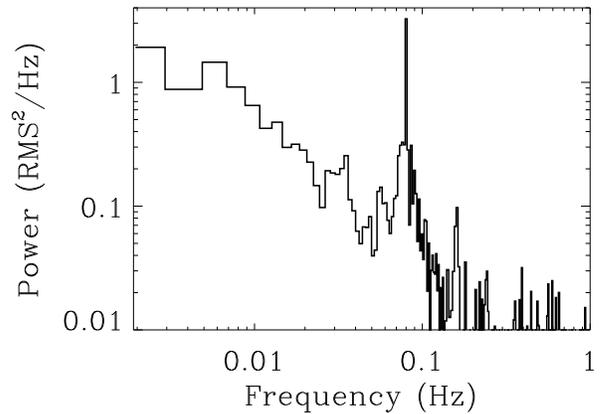,width=\columnwidth}
\caption{0.00195--1~Hz white-noise corrected power spectral density distribution of 
\xte\ Obs. 1 \& 3 \correc{obtained from the 2--20.2~keV high-resolution light curve.}}
\label{fig:psd}
\end{figure}

Although the first hint for the presence of a pulsation came from \swift/XRT 
data \citep{2009ATel.1998....1R}, we focus here only on the data from \xte/PCA 
which is the instrument dedicated to the timing analysis of astrophysical sources.
Also, we only consider here Obs. X1 and X3 which both have 2 PCUs on and exposure times 
longer than 2.5~ks, in order to increase the statistical significance of the pulse. 
We produced power spectral density (PSD) distributions between 0.00195 
and 1024~Hz with {\tt powspec v1.0} averaging all sub-intervals of 512~s 
long of each of the two observations.  The continuum of the 0.00195--1024~Hz 
PSD is equally well represented by a model consisting 
of a constant (to account for the contribution of Poisson noise), and either a power law 
(of index $\Gamma=-1.35\pm0.08$) 
or a zero-centred Lorentzian (with a width of $1.6\pm0.4\times 10^{-2}$~Hz). 
Fig. \ref{fig:psd} displays the white-noise corrected PSD restricted to the 
0.00195--1~Hz range. Above  $\sim0.5$~Hz, the PSD is compatible with 
white noise.
A strong peak at around 0.08~Hz and its first harmonic at about 0.16 Hz 
can be seen above the PSD continuum (Fig.~\ref{fig:psd}). The main 
peak (at 0.08~Hz) is contained within 
a single frequency bin which is indicative of a coherent pulsation. However, we point out  
that the bottom of this thin peak is broadened.  This broad feature can be fitted with a Lorentzian 
 of centroid frequency $0.078\pm0.002$~Hz, and a width of $0.020\pm0.008$~Hz. 
It has a (raw) power of 11.1\% RMS. We also note the presence of a possible  
feature at $\sim0.035$~Hz, yet the inclusion of a second Lorentzian to the model does not 
improve the fit in a significant way. To study the properties of the coherent pulsation, 
we barycentred the 1-s light curves 
with {\tt barycorr v1.8} using the orbital ephemeris of the satellite, and we corrected the light curves for the background. The pulsation period was searched for using a Lomb-Scargle periodogram 
as described in \citet{press89} with the errors calculated from the periodogram 
following \citet{horne86} \citep[see also the original work of][]{kovacs81}.  
This is the same method used by \citet{rodrigue06_1632} to 
find the pulse's most precise period in \object{IGR J16320$-$4751}. 
The light curves were folded with {\tt efold v1.1}.
\begin{figure}
 \epsfig{file=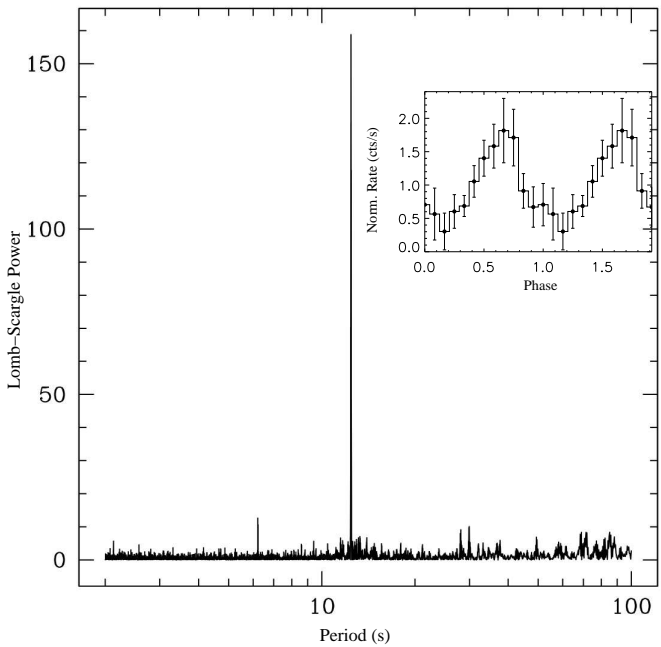,width=\columnwidth}
\caption{Lomb-Scargle periodogram of \xte\ Obs. 1 \& 3. The prominent peak is found 
at a period of $\sim12.44$~s. The insert shows the light curve folded at 12.44~s.}
\label{fig:PSD}
\end{figure}
We combined Obs. X1 \& X3 to obtain the most precise measure of the pulsation. A very 
prominent peak is visible in the periodogram (Fig.~\ref{fig:PSD}), together with its first 
harmonic.  We derived a pulse period of P=12.43781(3)~s (0.080400~Hz).
The light curve, folded at a period of 12.44~s, is shown in Fig.~\ref{fig:PSD} 
as an insert. 
The pulse fraction was calculated in a standard way 
\citep[$P=\frac{I_{max}-I_{min}}{I_{max}+I_{min}}$, see e.g.][]{rodrigue06_1632}.
Without taking the emission of the Galactic ridge into account, we estimate 
a pulse fraction $P=28\pm2.3\%$. Given the source's location in the Galactic plane which 
has a non-negligible contribution to the total flux, this value underestimates 
the true pulse fraction, and should be considered as a lower limit. 
According to the results of the spectral analysis
presented in the previous section, we obtained the ``true'' background following 
Eq. \ref{eq:bgd}, which includes the estimated contribution from the Galactic 
ridge at the position of \igr. The light curves corrected for this total background yields $P=40\pm4.4\%$. 

\section{Discussion and conclusions}
\begin{figure}
\epsfig{file=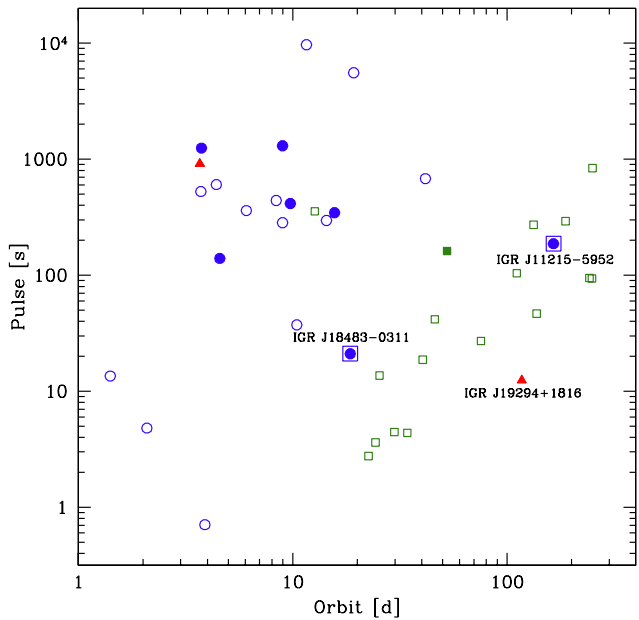,width=\columnwidth}
\caption{Most recent version of the ``Corbet diagram'' including all sources 
detected with \integ\ (Bodaghee et al. 2009 in prep.). Squares represent Be-HMXBs, 
circles Supergiant-HMXBs (Sg-HMXB), 
circles in squares are SFXTs, filled symbols represent sources discovered by
\integ. The triangles are HMXBs of unknown nature (including \igr). The positions of 
\igr\ and of the two SFXTs lying in the Be region of the plot are highlighted. }
\label{fig:corbet}
\end{figure}

We have reported here the results obtained from a multi-instrumental 
campaign dedicated to the X-ray properties of new \integ\ sources.
The refinement of the X-ray position provided by the \swift/XRT observations enabled us to identify a possible counterpart at infrared wavelengths. The 
differences of dereddened magnitudes (Table \ref{tab:mag}) do not lead to 
any of the spectral types tabulated in \citet{tokunaga00}. With 
\nh=2.0$\times 10^{22}$ cm$^{-2}$,  J$-$H=0.28 would indicate an F7 V or F8 I 
star. However the value of H$-$K$_{\mathrm s}$ is inconsistent with both possibilities. 
 With \nh=2.73$\times 10^{22}$ cm$^{-2}$, J$-$H seems too high for any spectral type, although
we remark a marginal compatibility (at the edge of the errors on the magnitude) with an 
O9.5 V star \citep[J$-$H=$-0.13$, H$-$K$_{\mathrm s}$=$-$0.04][]{tokunaga00}.  This 
value is, however, a measure of the interstellar absorption through the whole Galaxy. Since \igr\ 
probably lies closer than the other end of the Galaxy, it is 
likely that the interstellar absorption along the line of sight is lower. In particular, we note 
that with \nh=2.5$\times 10^{22}$ cm$^{-2}$, we obtain J$-$H=$-0.04$ and 
H$-$K$_{\mathrm s}$=0.05, which is very close to the values tabulated for a B3 I star 
\citep[J$-$H=$-0.03$, H$-$K$_{\mathrm s}$=0.03][]{tokunaga00}. In that case, the source
would lie at a distance $d \gtrsim 8$ kpc.\\
\indent The X-ray behaviour of the source is indicative of an HMXB:
the X-ray spectra are power law like in shape with a hard photon index. 
One of the spectra (Obs. S1 with the longest exposure) 
shows evidence for absorption in clear excess to the 
absorption on the line of sight, which may indicate some intrinsic absorption in 
this source. However, the evidence is marginal in the other spectra, which could suggest that the intrinsic absorption varies in this 
system, as has been observed in a number of HMXBs. 
A long-term periodicity was revealed in the \swift/BAT data which confirms the binarity of the 
source \citep{2009ATel.2008....1C}. We clearly detect an X-ray pulse at a period of 
12.44~s, indicating the presence of an X-ray pulsar. We estimate a pulse fraction 
$P=40\%$. Apart from millisecond X-ray pulsars, found in systems that are at the end of 
the evolutionary path of X-ray binaries (and are LMXBs), pulsars are young objects 
that are usually found in HMXBs. We also remark that the PSD of \igr\ shows a
broadening at the bottom of the coherent pulsation.  Sidelobes and other noise features
around coherent pulsation signals have been seen in other HMXBs 
\citep[e.g.][in, respectively, \object{4U 1626$-$67}, and 
\object{XTE J0111.2$-$7317}]{kommers98,kaur07}. 
Such features can be produced by artificial effects (such as the finite
length of the time intervals used to make the PSD), or they can be
real if, e.g., a quasi-periodic oscillation (QPO) beats with the coherent 
signal \citep{kommers98,kaur07}. In the case of \igr, a possible QPO might be present 
at $\sim0.035$~Hz (Fig. \ref{fig:psd}), even if the quality of the data does 
not allow us to firmly establish its presence.  On the other hand, 
we cannot exclude that the $0.078$~Hz feature is itself a QPO at a frequency 
close to that of the coherent pulsation. Given the faintness of the source, 
we cannot conclude further on that matter.  \\
\indent In fact, \igr\  lies in a region populated 
by Be-HMXBs in the so-called ``Corbet diagram'' \citep{corbet86,2009ATel.2008....1C}, as 
demonstrated in Fig. \ref{fig:corbet}. Note that this plot is the most recent update of the Corbet 
diagram for \integ\ sources as of June, 2009 (Bodaghee et al. 2009 in prep.). 
Recently, \citet{arash07} further explored the 
parameter spaces occupied by high-energy sources. They noticed 
that Be-HMXB and supergiant HMXB also segregate in different parts of the 
\nh\ vs. orbital period and \nh\ vs. spin period diagrams.  The value of the 
absorption we obtained through our spectral analysis, combined with the values of 
the spin and orbital periods, make \igr\ lie in a region populated by Be-HMXBs in these
diagrams as well. At first order, one can easily understand the 
segregation in these three different diagrams as results of the age, type of accretion, 
and probable type of orbit. Be systems are younger, and hence have eccentric orbits, 
longer orbital periods, and shorter spin periods, whereas supergiant systems are older, are 
mostly circularised with shorter orbital periods, and longer spin periods. In these latter systems, 
in addition, the compact object is embedded in the wind of the companion (the feed matter for
accretion) which explains their (usually) higher intrinsic absorption.  In this respect, 
all our results point towards \igr\ being a Be-HMXB \citep[the same conclusion, although only based 
 on the Corbet diagram, was presented by][]{2009ATel.2008....1C} .\\
\indent \integ\ has unveiled a new (sub-)type of supergiant HMXB, characterised by 
short ($\sim$hours) and intense flares seen at X-ray energies: the so-called SFXTs. 
Models of SFXTs 
involve stochastic accretion of clumps from the (heterogeneous) wind of the 
supergiant \citep[e.g.][]{zand05, negueruela06}, on top
of longer, but fainter, periods of activity \citep[e.g.][]{sidoli07}. 
In this respect, the detection of periods of short and intense 
flares in \igr\ with \integ\ (Fig. \ref{fig:igrlite}), a behaviour typical of SFXTs, 
raises the possibility that this source belongs 
to this new class. Its position in the Corbet and the \nh\ vs. spin or orbital period 
diagrams (Fig.~\ref{fig:corbet}) is not a definitive 
contradiction to this hypothesis, since systematic X-ray studies of SFXTs
have shown that, contrary to other HMXBs, they 
seem to populate any part of the diagrams. In particular, \object{IGR J18483$-$0311}  
and \object{IGR J11215$-$5952}, the only two SFXTs for which both orbital and pulse
periods are known, lie in the region of Be-systems in these representations 
(Fig. \ref{fig:corbet}). The former source has a pulse period 
of 21.05~s, an orbital period of 18.5~d, and has a B0.5 Ia companion 
\citep[see][and references therein]{rahoui08_18483}. The latter has a pulse period of about 
187~s, an orbital period of about 165~d \citep[e.g.][]{sidoli07, ducci09}, and is associated
with a B supergiant \citep{negueruela05}. To reconcile the fact that they are long period systems
 with a rather low absorption, an eccentric orbit is invoked. \igr\ could be the third member 
of SFXTs (with a known pulse period) having a long and eccentric orbit. 
In this case, this may point towards 
the existence of an evolutionary link between Be-HMXBs and eccentric 
SFXTs (Liu et al. 2009 in prep.). 
The quality of the data is not good enough to permit us to make any firm conclusions concerning the nature of the system, and only an identification of the optical counterpart's spectral type will resolve this issue.

\begin{acknowledgements}
JR acknowledges useful and productive discussions with Farid Rahoui. We thank the referee 
for the careful reading of this paper and his/her comments that helped us to improve it. We 
would like to warmly thank the RXTE and Swift PIs and mission planners for 
having accepted to perform these observations, and the efficient scheduling in 
simultaneity with one of the INTEGRAL pointings. AP 
acknowledges the Italian Space Agency financial
  support via contract I/008/07/0. JAZH acknowledges the 
Swiss National Science Foundation for financial support.
   This research has made use of the USNOFS Image and Catalogue Archive
operated by the United States Naval Observatory, Flagstaff Station
(http://www.nofs.navy.mil/data/fchpix/)
 This research has made use of the SIMBAD database, operated at CDS, Strasbourg, France.
It also makes use of data products from the Two Micron All Sky Survey, which 
is a joint project of the University of Massachusetts and the Infrared Processing 
and Analysis Center/California Institute of Technology, funded by the National 
Aeronautics and Space Administration and the National Science Foundation.
The Digitized Sky Surveys were produced at the Space Telescope Science Institute 
under U.S. Government grant NAG W-2166. The images of these surveys are based 
on photographic data obtained using the Oschin Schmidt Telescope on Palomar Mountain 
and the UK Schmidt Telescope. The plates were processed into the present compressed 
digital form with the permission of these institutions.
The Second Palomar Observatory Sky Survey (POSS-II) was made by the California Institute 
of Technology with funds from the National Science Foundation, the National Geographic 
Society, the Sloan Foundation, the Samuel Oschin Foundation, and the Eastman Kodak 
Corporation. 
\end{acknowledgements}


\begin{thebibliography}{30}
\expandafter\ifx\csname natexlab\endcsname\relax\def\natexlab#1{#1}\fi

\bibitem[{{Bodaghee} {et~al.}(2007){Bodaghee}, {Courvoisier}, {Rodriguez},
  {Beckmann}, {Produit}, {Hannikainen}, {Kuulkers}, {Willis}, \&
  {Wendt}}]{arash07}
{Bodaghee}, A., {Courvoisier}, T.~J.-L., {Rodriguez}, J., {et~al.} 2007, \aap,
  467, 585

\bibitem[{{Cardelli} {et~al.}(1989){Cardelli}, {Clayton}, \&
  {Mathis}}]{cardelli89}
{Cardelli}, J.~A., {Clayton}, G.~C., \& {Mathis}, J.~S. 1989, \apj, 345, 245

\bibitem[{{Condon} {et~al.}(1998){Condon}, {Cotton}, {Greisen}, {Yin},
  {Perley}, {Taylor}, \& {Broderick}}]{condon98}
{Condon}, J.~J., {Cotton}, W.~D., {Greisen}, E.~W., {et~al.} 1998, \aj, 115,
  1693

\bibitem[{{Corbet}(1986)}]{corbet86}
{Corbet}, R.~H.~D. 1986, \mnras, 220, 1047

\bibitem[{{Corbet} \& {Krimm}(2009)}]{2009ATel.2008....1C}
{Corbet}, R.~H.~D. \& {Krimm}, H.~A. 2009, The Astronomer's Telegram, 2008, 1

\bibitem[{{Dame} {et~al.}(2001){Dame}, {Hartmann}, \& {Thaddeus}}]{dame01}
{Dame}, T.~M., {Hartmann}, D., \& {Thaddeus}, P. 2001, \apj, 547, 792

\bibitem[{{Ducci} {et~al.}(2009){Ducci}, {Sidoli}, {Mereghetti}, {Paizis}, \&
  {Romano}}]{ducci09}
{Ducci}, L., {Sidoli}, L., {Mereghetti}, S., {Paizis}, A., \& {Romano}, P.
  2009, \mnras, 398, 2152

\bibitem[{{Horne} \& {Baliunas}(1986)}]{horne86}
{Horne}, J.~H. \& {Baliunas}, S.~L. 1986, \apj, 302, 757

\bibitem[{{in't Zand}(2005)}]{zand05}
{in't Zand}, J.~J.~M. 2005, \aap, 441, L1

\bibitem[{{Kalberla} {et~al.}(2005){Kalberla}, {Burton}, {Hartmann}, {Arnal},
  {Bajaja}, {Morras}, \& {P{\"o}ppel}}]{kalberla05}
{Kalberla}, P.~M.~W., {Burton}, W.~B., {Hartmann}, D., {et~al.} 2005, \aap,
  440, 775

\bibitem[{{Kaur} {et~al.}(2007){Kaur}, {Paul}, {Raichur}, \& {Sagar}}]{kaur07}
{Kaur}, R., {Paul}, B., {Raichur}, H., \& {Sagar}, R. 2007, \apj, 660, 1409

\bibitem[{{Kommers} {et~al.}(1998){Kommers}, {Chakrabarty}, \&
  {Lewin}}]{kommers98}
{Kommers}, J.~M., {Chakrabarty}, D., \& {Lewin}, W.~H.~G. 1998, \apjl, 497,
  L33+

\bibitem[{{Kovacs}(1981)}]{kovacs81}
{Kovacs}, G. 1981, \apss, 78, 175

\bibitem[{{Negueruela} {et~al.}(2005){Negueruela}, {Smith}, \&
  {Chaty}}]{negueruela05}
{Negueruela}, I., {Smith}, D.~M., \& {Chaty}, S. 2005, The Astronomer's
  Telegram, 470, 1

\bibitem[{{Negueruela} {et~al.}(2006){Negueruela}, {Smith}, {Reig}, {Chaty}, \&
  {Torrej{\'o}n}}]{negueruela06}
{Negueruela}, I., {Smith}, D.~M., {Reig}, P., {Chaty}, S., \& {Torrej{\'o}n},
  J.~M. 2006, in ESA Special Publication, Vol. 604, The X-ray Universe 2005,
  ed. A.~{Wilson}, 165--+

\bibitem[{{Prat} {et~al.}(2008){Prat}, {Rodriguez}, {Hannikainen}, \&
  {Shaw}}]{lionel08}
{Prat}, L., {Rodriguez}, J., {Hannikainen}, D.~C., \& {Shaw}, S.~E. 2008,
  \mnras, 389, 301

\bibitem[{{Predehl} \& {Schmitt}(1995)}]{predehl95}
{Predehl}, P. \& {Schmitt}, J.~H.~M.~M. 1995, \aap, 293, 889

\bibitem[{{Press} \& {Rybicki}(1989)}]{press89}
{Press}, W.~H. \& {Rybicki}, G.~B. 1989, \apj, 338, 277

\bibitem[{{Rahoui} \& {Chaty}(2008)}]{rahoui08_18483}
{Rahoui}, F. \& {Chaty}, S. 2008, \aap, 492, 163

\bibitem[{{Rahoui} {et~al.}(2008){Rahoui}, {Chaty}, {Lagage}, \&
  {Pantin}}]{rahoui08}
{Rahoui}, F., {Chaty}, S., {Lagage}, P.-O., \& {Pantin}, E. 2008, \aap, 484,
  801

\bibitem[{{Rodriguez} {et~al.}(2006){Rodriguez}, {Bodaghee}, {Kaaret},
  {Tomsick}, {Kuulkers}, {Malaguti}, {Petrucci}, {Cabanac}, {Chernyakova},
  {Corbel}, {Deluit}, {di Cocco}, {Ebisawa}, {Goldwurm}, {Henri}, {Lebrun},
  {Paizis}, {Walter}, \& {Foschini}}]{rodrigue06_1632}
{Rodriguez}, J., {Bodaghee}, A., {Kaaret}, P., {et~al.} 2006, \mnras, 366, 274

\bibitem[{{Rodriguez} {et~al.}(2008{\natexlab{a}}){Rodriguez}, {Hannikainen},
  {Shaw}, {Pooley}, {Corbel}, {Tagger}, {Mirabel}, {Belloni}, {Cabanac},
  {Cadolle Bel}, {Chenevez}, {Kretschmar}, {Lehto}, {Paizis}, {Varni{\`e}re},
  \& {Vilhu}}]{rodrigue08_1915a}
{Rodriguez}, J., {Hannikainen}, D.~C., {Shaw}, S.~E., {et~al.}
  2008{\natexlab{a}}, \apj, 675, 1436

\bibitem[{{Rodriguez} {et~al.}(2008{\natexlab{b}}){Rodriguez}, {Tomsick}, \&
  {Chaty}}]{rodrigue08_igr}
{Rodriguez}, J., {Tomsick}, J.~A., \& {Chaty}, S. 2008{\natexlab{b}}, \aap,
  482, 731

\bibitem[{{Rodriguez} {et~al.}(2009){Rodriguez}, {Tuerler}, {Chaty}, \&
  {Tomsick}}]{2009ATel.1998....1R}
{Rodriguez}, J., {Tuerler}, M., {Chaty}, S., \& {Tomsick}, J.~A. 2009, The
  Astronomer's Telegram, 1998, 1

\bibitem[{{Sidoli} {et~al.}(2007){Sidoli}, {Romano}, {Mereghetti}, {Paizis},
  {Vercellone}, {Mangano}, \& {G{\"o}tz}}]{sidoli07}
{Sidoli}, L., {Romano}, P., {Mereghetti}, S., {et~al.} 2007, \aap, 476, 1307

\bibitem[{{Skrutskie} {et~al.}(2006){Skrutskie}, {Cutri}, {Stiening},
  {Weinberg}, {Schneider}, {Carpenter}, {Beichman}, {Capps}, {Chester},
  {Elias}, {Huchra}, {Liebert}, {Lonsdale}, {Monet}, {Price}, {Seitzer},
  {Jarrett}, {Kirkpatrick}, {Gizis}, {Howard}, {Evans}, {Fowler}, {Fullmer},
  {Hurt}, {Light}, {Kopan}, {Marsh}, {McCallon}, {Tam}, {Van Dyk}, \&
  {Wheelock}}]{skrutskie06}
{Skrutskie}, M.~F., {Cutri}, R.~M., {Stiening}, R., {et~al.} 2006, \aj, 131,
  1163

\bibitem[{{Strohmayer} {et~al.}(2009){Strohmayer}, {Rodriguez}, {Markwardt},
  {Tomsick}, {Bodaghee}, {Chaty}, {Corbel}, \& {Paizis}}]{2009ATel.2002....1S}
{Strohmayer}, T., {Rodriguez}, J., {Markwardt}, C., {et~al.} 2009, The
  Astronomer's Telegram, 2002, 1

\bibitem[{{Tokunaga}(2000)}]{tokunaga00}
{Tokunaga}, A.~T. 2000, {Infrared Astronomy}, ed. A.~N. {Cox}, 143--+

\bibitem[{{Tuerler} {et~al.}(2009){Tuerler}, {Rodriguez}, \&
  {Ferrigno}}]{2009ATel.1997....1T}
{Tuerler}, M., {Rodriguez}, J., \& {Ferrigno}, C. 2009, The Astronomer's
  Telegram, 1997, 1

\bibitem[{{Valinia} \& {Marshall}(1998)}]{valinia98}
{Valinia}, A. \& {Marshall}, F.~E. 1998, \apj, 505, 134

\end{thebibliography}
\end{document}